\documentclass[11pt]{article} 
\usepackage{graphicx}
\usepackage{bm}
\usepackage{epsfig} 
\usepackage{amsmath}
\usepackage{fullpage}
\usepackage[round,numbers,sort&compress]{natbib} 
\usepackage{endfloat}
\begin{document} 

\addtolength{\textheight}{0.5in}

\title{Entropic elasticity of DNA with a permanent kink} 
\author{Jinyu
  Li$^1$, Philip C. Nelson$^2$, and M. D. Betterton$^{3*}$\\
 $^1$Department of Applied Mathematics,
  University of Colorado at Boulder, \\ Boulder, CO, jinyu.li@colorado.edu\\
$^2$ Department of Physics and Astronomy, 
  University of Pennsylvania, \\ Philadelphia, PA, pcn@physics.upenn.edu\\
$^3$Department of Physics,
  University of Colorado at Boulder, \\ Boulder, CO, mdb@colorado.edu \\
$^*$Corresponding author. Address: Department of Physics, University
  of Colorado at Boulder,\\ 390 UCB, Boulder, CO~80309 USA}

\maketitle

\begin{abstract}
  Many proteins interact with and deform double-stranded DNA in cells.
  Single-molecule experiments have studied the elasticity of DNA with
  helix-deforming proteins, including proteins that bend DNA. These
  experiments increase the need for theories of DNA elasticity which
  include helix-deforming proteins. Previous theoretical work on bent
  DNA has examined a long DNA molecule with many nonspecifically
  binding proteins. However, recent experiments used relatively short
  DNA molecules with a single, well-defined bend site. Here we develop
  a simple, 
 theoretical description of the
  effect of a single bend.  We then include the description of the
  bend in the finite worm like chain model (FWLC) of short DNA
  molecules attached to beads. We predict how the DNA force-extension
  relation changes due to formation of a single permanent kink, at all
  values of the applied stretching force. Our predictions show that
  high-resolution single-molecule experiments could determine the bend
  angle induced upon protein binding.
\end{abstract}


\section{Introduction}

In cells, many different interactions between DNA and proteins occur,
processes which are essential to gene expression, genome replication,
and cellular DNA management.  One major class of proteins interacts
with DNA and mechanically deforms the double helix by wrapping,
looping, twisting, or bending DNA \cite{dicker98,luscom00}.  Examples
include DNA-packaging proteins and transcription factors which
regulate gene expression.  The mechanical deformation of the DNA may
be important for gene expression: it has been suggested that DNA
deformation by transcription factors may help other proteins bind to
the DNA and initiate transcription.

\begin{figure} 
   \begin{center}
      \includegraphics*[width=2.5in]{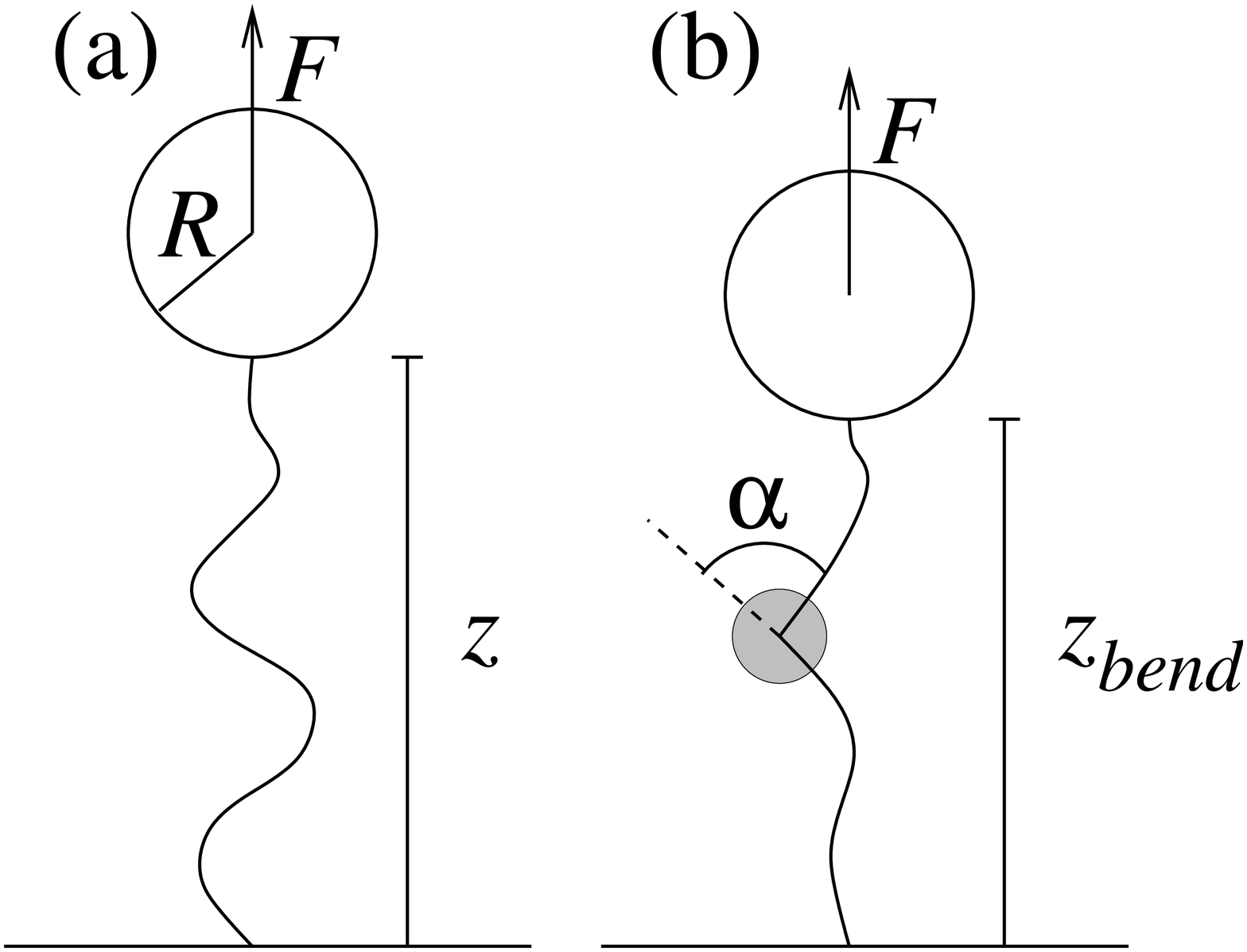}
      \caption{Typical experimental geometry of single-molecule force
        microscopy measurements. The DNA molecule is attached at one
        end to a surface and at the other end to a bead (radius $R$).
        A force $F$ is applied to the bead. (a) DNA molecule in the
        absence of bound protein.  The mean end-to-end extension is $
        z $.  (b) DNA molecule with a single bend-inducing protein
        bound.  The protein bends the DNA backbone through the
        external angle $\alpha$ at the bend site.  As a result, the
        mean extension decreases to $ z_{bend}$.}
      \label{bendgeom}
   \end{center} 
\end{figure}

The deformation of DNA by proteins can be detected in single-molecule
force microscopy. In this experimental method, force is applied to
individual DNA molecules and the DNA end-to-end extension is measured
(figure \ref{bendgeom}). Single-molecule force microscopy has been
used to detect the deformation of DNA caused by protein binding
\cite{noort04,skoko04,yan04a,broek05,dixit05,mccaul05}.  In these
experiments the DNA end-to-end extension changes when a
deformation-inducing protein binds.  Varying the applied force allows
one to probe the deformation and better understand the details of the
protein-DNA interaction.

In this paper we focus on proteins that bend the DNA backbone and
develop theoretical predictions of the force-extension behavior of
bent DNA.  Our description is based on the worm-like chain theory
(WLC) \cite{bustam94,marko95,bouch99}. The WLC predicts the average
end-to-end extension $z$ of a semiflexible polymer, given the force
$F$ applied to the ends of the chain and the values of two constant
parameters (the contour length $L$ and the persistence length $A$).
However, DNA elastic behavior is altered by backbone-deforming
proteins, an effect that is not included in the traditional WLC.
Extended theories have been developed which combine the WLC treatment
of DNA elasticity with local bends.  Rivetti \textit{et al.}\ 
addressed the case of zero applied force \cite{rivet98}, while Yan and
Marko have described the changes in the force-extension behavior of a
long polymer to which many kink-inducing proteins can bind
nonspecifically \cite{yan03}.  Similarly, Popov and Tkachenko studied
the effects of a large number of reversible kinks \cite{popov05},
Metzler \textit{et al.}\ studied loops formed by slip-rings
\cite{metz02a} , and Kuli\'c \textit{et al.} studied the high-force
limit of a kinked polymer \cite{kulic05}.

Previous theoretical work has focused on large numbers of reversible
kinks or the limit of low or high applied force. However, recent
single-molecule experiments have examined relatively short DNA
molecules with a single specific kink site, over a range of applied
force \cite{dixit05}. Therefore a theory is needed which applies to
({\it i}) one kink site and ({\it ii}) a polymer of finite contour
length ($L/A \sim 1-10$).  Recently, we introduced a modified solution
of the WLC applicable to polymers of this length, and demonstrated
that applying the traditional WLC solution to molecules with $L/A \sim
1-10$ can lead to significant errors \cite{li05}. Our finite worm-like
chain solution (FWLC) includes both finite-length effects, often
neglected in WLC calculations, and the effect of the rotational
fluctuations of a bead attached to the end of the chain.

Here we formulate a theoretical description of a single kink induced
by a protein, and extend the FWLC treatment to include such local
distortions.  Our theory has a simple analytical formulation for the
case of a force-independent bend angle, i.e., a rigid protein-DNA
complex. Our predictions are relevant to experiments like those of
Dixit \textit{et al.}\ \cite{dixit05}, which detect with high
resolution a single bend induced in a relatively short DNA molecule.
Although we will primarily focus on the case of a single bend angle,
our method can also describe a kink which takes on different angles
with different probabilities. This model could be relevant to a
binding protein that can fluctuate between different binding
conformations with different kink angles \cite{parkh01}.

\section{Theory}

\begin{figure} 
   \begin{center}
      \includegraphics*[width=3in]{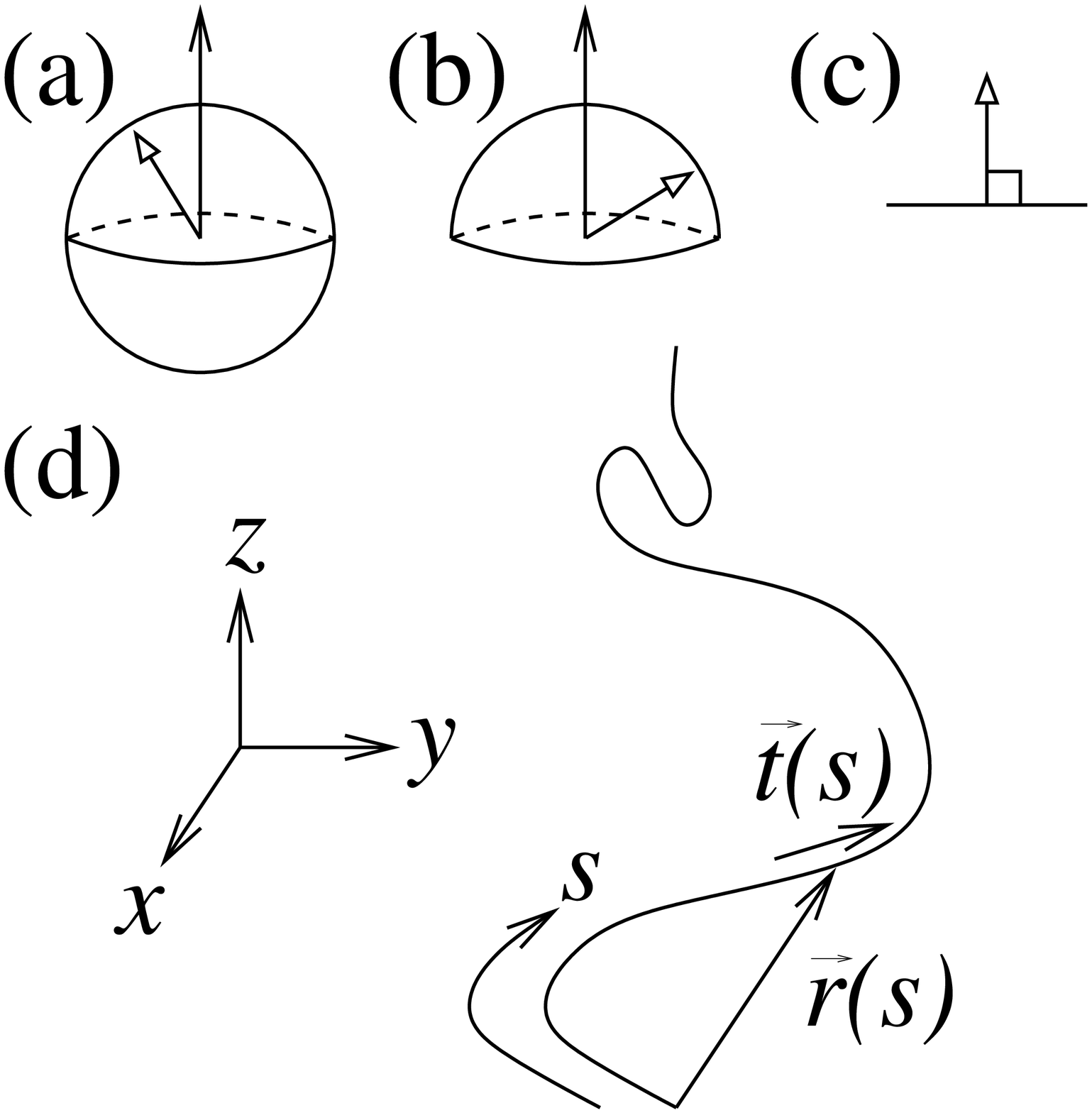}
      \caption{Boundary conditions and coordinates. (a)
        Unconstrained. (b) Half-constrained. (c) Normal. (d)
        Coordinates.}
      \label{bc-coord}
   \end{center} 
\end{figure} 

\subsection{FWLC theory of unkinked DNA}

The classic WLC model \cite{marko95} and the FWLC theory \cite{li05},
which includes finite-length effects, describe an inextensible polymer
with isotropic bending rigidity.  The bending rigidity is
characterized by the persistence length, $A$, the length scale over
which thermal fluctuations randomize the chain orientation. We assume
that the twist is unconstrained and can be neglected (as is the case,
for example, in optical tweezer experiments).

The chain energy function includes terms which represent the bending
energy and the work done by the applied force:
\begin{equation} 
E=\int_0^\ell ds \left( \frac{\kappa^2}{2} - f {\bf \hat{\bf z}}\cdot{\bf
      \hat{t}}  \right),
\end{equation}
where $E$ is the energy divided by the thermal energy $k_BT$,
$\ell=L/A$, $s$ denotes arc length divided by the persistence length
$A$, and all other lengths are similarly measured in units of $A$. The
quantity $f$ is the force multiplied by $A/k_BT$, and we assume the
force is applied in the $\hat{\bf z}$ direction. The total extension
of the chain is $z=\int ds\ {\bf \hat{\bf z}}\cdot{\bf \hat{t}}$. The
curvature $\kappa$ can be defined in terms of arc-length derivatives
of the chain coordinate (figure \ref{bc-coord}d). If the chain
conformation is described by a space curve ${\bf r}(s)$ and the unit
vector tangent to the chain is ${\bf \hat{t}}(s)$, then $\kappa =
\left| \frac{\partial^2 {\bf r}}{\partial s^2}\right|= \left|
  \frac{\partial {\bf \hat{t}}}{\partial s}\right|$.

The chain partition function weights contributions from different
polymer conformations \cite{fixman73,yamak76}.  If the ends of the
chain are held at fixed orientations, we have
\begin{equation} 
Z({\bf \hat{t}}_f,\ell;{\bf \hat{t}}_i,0)
=\int D{\bf \hat{t}} \ \exp \left[- \int_0^\ell ds \left( \frac{1
      }{2}(\partial_s{\bf \hat{t}})^2 - f {\bf \hat{\bf z}}\cdot{\bf
      \hat{t}} \right)\right],
\end{equation} 
where the integral in $D{\bf \hat{t}}$ is over all possible paths
between the two endpoints of the chain with the specified
orientations.  The partition function can be interpreted as a
propagator which connects the probability distribution for the tangent
vector at point $s$, $\psi({\bf \hat{t}},s)$ to the same probability
distribution at point $s'$:
\begin{equation}
\psi({\bf \hat{t}},s) = \int d{\bf \hat{t}}' \ Z({\bf \hat{t}},s;{\bf
  \hat{t}}',s')\ \psi({\bf \hat{t}}',s').
\label{propagator} 
\end{equation}
From this relation, one can derive a Schr\"odinger-like equation, which
describes the $s$ evolution of $\psi$  \cite{marko95}:
\begin{equation} 
\frac{\partial \psi}{\partial s}=\left(\frac{\nabla^2}{2}+f \cos
  \theta \right) \psi.
\label{hamilt}
\end{equation}
Here $\nabla^2$ is the two-dimensional Laplacian on the surface of the
unit sphere and $\cos \theta= \hat{\bf z} \cdot \hat{\bf t}$.

For relatively short DNA molecules ($\ell \sim 1-10$), the boundary
conditions at the ends of the chain \cite{li05,samuel02} and
bead rotational fluctuations become important.  The boundary
conditions are specified by two probability density functions,
$\psi({\bf \hat{t}},s=0)$ and $\psi({\bf \hat{t}},s=L)$.  The boundary
conditions modify the force-extension relation, and enter the full
partition function matrix element via
\begin{equation}
  \label{fullpart}
  Z_{tot}=\int d{\bf \hat{t}}_i \ d {\bf \hat{t}}_f \ \psi({\bf
    \hat{t}}_i,0) Z({\bf \hat{t}}_i,0;{\bf \hat{t}}_f,L) \psi({\bf
    \hat{t}}_f,L).
\end{equation}
Rotational fluctuations of the bead(s) attached to the end of the DNA
complicate the analysis of experiments.
What is observed and controlled is not the endpoint of the (invisible)
DNA chain, but rather the bead's center. The relation between these
distinct points fluctuates as the bead performs rotational Brownian
motion. The FLWC theory accounts for these fluctuations via
an effective boundary condition at the end(s) of the chain, which
depends on applied force, bead radius, and the nature of the link
joining the bead to the polymer \cite{li05}. We will study boundary
conditions that are azimuthally symmetric; thus our end boundary
conditions will be functions of $\hat{\bf t}\cdot \hat{\bf z}$ only.

\subsection{Fixed-angle bend}

We now suppose that our chain contains a permanent bend, whose
location along the DNA, and angle, are fixed, independent of applied
force. In this paper we will also neglect
force-induced unbinding of the deforming protein. (These effects are
straightforward to incorporate into our analysis.) In addition, we
neglect twist stiffness, which is legitimate since we wish to study a
single bend in a polymer with unconstrained twist.  (Twist stiffness
effects will be important for experiments in which multiple bends
occur or twist is constrained.)

The Schr\"odinger-like equation \eqref{hamilt} must be modified by the
inclusion of a ``bend operator'' which transforms $\psi$ at the bend.
Suppose that the kink occurs at position $s_o$.  Given the
tangent-vector probability distribution $\psi$ at $s_o-\epsilon$
(where $\epsilon$ is infinitesimal), our goal is to determine
$\psi({\bf \hat{t}},s_o+\epsilon)$, the distribution just after the kink. If we
denote the exterior angle of the kink by $\alpha$, then ${\bf
  \hat{t}}_{s_o-\epsilon} \cdot {\bf \hat{t}}_{s_o+\epsilon}=\cos
\alpha$. Because twist is unconstrained, we may average over
rotations; effectively, the bend occurs with uniform probability in
the azimuthal angle: if ${\bf \hat{t}}_{s_o-\epsilon}$ points directly
along the $\hat{\bf z}$-axis, then ${\bf \hat{t}}_{s_o+\epsilon}$ is
uniformly distributed in a cone at angle $\alpha$ to the $\hat{\bf
  z}$-axis.

The bend operator then can be written using the kernel
\begin{equation}
K_{\alpha}({\bf \hat{t}},{\bf \hat{t}}')=\frac{1}{2 \pi} \delta({\bf
  \hat{t}}\cdot {\bf \hat{t}}'-\cos \alpha).
\label{bendop}
\end{equation}
The probability distribution $\psi({\bf \hat{t}}',s_o-\epsilon)$ of
tangent-vector angles just before the kink is related to the
distribution $\psi({\bf \hat{t}},s_o+\epsilon)$ just after the kink by
\begin{equation} 
\psi({\bf \hat{t}},s_o+\epsilon)= \int d {\bf \hat{t}}'\
K_{\alpha}({\bf \hat{t}},{\bf \hat{t}}') \ \psi({\bf
  \hat{t}}',s_o-\epsilon).
\label{psitrans}
\end{equation}
Below (section \ref{bendopcalc}) we show that spherical harmonics
diagonalize the operator \eqref{bendop}.

\subsubsection{Distribution of bend angles}

Suppose that the bend occurs not for a single fixed angle, but a
distribution of angles. We assume that ${\bf \hat{t}}_{s_o-\epsilon}
\cdot {\bf \hat{t}}_{s_o+\epsilon}=\cos \alpha=u$ is distributed
according to the probability density function $h(u)$, where $h$ is
normalized so that $\int d \phi \int du \ h(u)=1$. Then the
bend-operator kernel can be written as an integral over the
probability distribution:
\begin{equation}
K_{h}({\bf \hat{t}},{\bf \hat{t}}')=\frac{1}{2 \pi} \int_{-1}^{1} du \
h(u)\ \delta({\bf   \hat{t}}\cdot {\bf \hat{t}}'-u). 
\label{bendopdist}
\end{equation}

\section{Calculation}

The main quantity of interest in single-molecule experiments is the
force-extension relation, which can be determined by solving equation
\eqref{hamilt} for the tangent-vector probability distribution
$\psi({\bf \hat{t}},s)$.  The Schr\"odinger-like equation is solved
using separation of variables in $s$ and ${\bf \hat{t}}$, where the
angular dependence is expanded in spherical harmonics \cite{marko95}.
\begin{equation} 
\psi({\bf \hat{t}},s) = \sum_{j=0}^{\infty} \Psi_j(s) Y_{j0}({\bf
  \hat{t}}).
\label{psiser}
\end{equation} 
(By azimuthal symmetry, only the $m=0$ terms will enter in our
formulae.) In the basis of spherical harmonics, the operator in
equation (\ref{hamilt}) is a symmetric tridiagonal matrix $H$ 
with diagonal terms
\begin{equation} 
H_{j,j}=-\frac{j(j+1)}{2},
\end{equation} 
and off-diagonal terms
\begin{equation} 
H_{j,j+1}=\frac{f(j+1)}{\sqrt{(2j+1)(2j+3)}}.
\end{equation}
The vector of coefficients at $s$ is ${\bf \Psi}(s)=e^{sH}{\bf
  \Psi}(0)$ \cite{marko95,li05}. This expression for $\psi({\bf
  \hat{t}},s)$ is exact if the infinite series of spherical harmonics
is used.

\subsection{Force-extension relation}

Given the boundary conditions ${\bf \Psi}(s=0)$ and ${\bf
  \Psi}(s=\ell)$, the partition function is 
\begin{eqnarray} 
Z&=&{\bf \Psi}^T(s=\ell)e^{\ell H}{\bf \Psi}(s=0),\\
&=&\sum_{j,k}  \Psi_j(s=\ell) [e^{\ell H}]_{jk} \Psi_k(s=0).
\label{zinnprod}
\end{eqnarray} 
The fractional extension of the chain is 
\begin{equation} 
\frac{z}{L}=\frac{1}{\ell}\frac{\partial \ln Z}{\partial f}.
\label{forcextf}
\end{equation} 
We work in the ensemble relevant to most experiments, where the
extension is determined for fixed applied force (different ensembles
are not equivalent for single finite-length molecules
\cite{dhar02,keller03,sinha05}).  Equation \eqref{forcextf} applies
for a chain of any length. However, we can show the structure of the
partition function more clearly by separating $\ln Z$ into two terms:
one representing an infinite chain and a finite-length correction
\cite{li05}.  Let $B=e^H$, denote by $\lambda_*$ the largest
eigenvalue of $B$, and let ${\cal B}=B/ \lambda_*$. Then ${\cal B}$
has eigenvalues with magnitude less than or equal to 1 and the
logarithm of the partition function can be written
\begin{equation}
  \ln Z=  \ell \ln \lambda_* + \ln [{\bf \Psi}^T(s=\ell){\cal B}^{\ell}
  {\bf \Psi}(s=0)]. 
\label{znew}
\end{equation}
Only the first term is considered in the usual WLC solution; the
second term is the finite-length correction \cite{li05}.
Equation \eqref{znew} is an exact expression for $\ln Z$ which is
difficult to evaluatae analytically.  We numerically calculate the
force-extension relation by using equation \eqref{znew} with the
series truncated after $N$ terms. This expression can be accurately
numerically calculated, and the truncation error determined by
comparing the results with different $N$.  Our calculations use $N=30$
unless otherwise specified.

\subsubsection{Boundary conditions and bead rotational fluctuations}

The boundary conditions at $s=0$ and $s=\ell$ affect the
force-extension relation, because they alter the partition function as
shown in equation \eqref{znew}.  Some experiments appear to implement
``half-constrained'' boundary conditions, where the polymer is
attached to a planar wall by a freely rotating attachment point, and
the wall is perpendicular to the direction of applied force
\cite{nelson05}. In this case the tangent vector at the end of the
chain can point in any direction on the hemisphere outside the
impenetrable surface (figure \ref{bc-coord}(b)). The effects of
different boundary conditions on the force-extension relation are
considered in detail in reference \cite{li05}.  In the
``unconstrained'' boundary condition the tangent vector at the end of
the chain is free to point in any direction on the sphere (in $4 \pi$
of solid angle, figure \ref{bc-coord}a). In this case $\psi(\hat{\bf
  t})$ is independent of $\cos \theta$ and ${\bf
  \Psi}=(1,0,\cdots,0)$.  In the ``half-constrained'' boundary
conditions (figure \ref{bc-coord}b), the tangent vector at the end of
the chain can point in any direction on the hemisphere outside the
impenetrable surface; then the leading coefficients of ${\bf \Psi}$
are 1, 0.8660, 0, -0.3307, 0, 0.2073, 0.  In the ``normal'' boundary
condition, the tangent vector at the end of the chain is parallel to
the $\hat{\bf z}$ axis, normal to the surface (figure
\ref{bc-coord}c).  Then the coefficients of ${\bf \Psi}$ are all equal
to 1 \cite{normnote}.

The FWLC formulation can also average over rotational fluctuations of
spherical bead(s) attached to one or both ends of the polymer chain.
The result is an effective boundary condition that depends on applied
force and bead radius \cite{li05}. Both the case of perpendicular wall
attachment and bead attachment generate boundary conditions that are
invariant under rotations about the $\hat{z}$ axis (the direction in
which force is applied), and hence give boundary states of the form
given in equation \eqref{psiser}.

\subsection{Bend operator}
\label{bendopcalc}

We wish to represent the bend operator (equation \eqref{psitrans}) in
terms of spherical harmonics; the operator is diagonal in this basis.
Denote $x={\bf \hat{t}}\cdot {\bf \hat{t}}'$ and note that any
function of $x$ with $-1\le x\le 1$ can be written as a series of
Legendre polynomials \cite{jackson}:
\begin{equation} 
K_{\alpha}(x)=\sum_{l=0}^{\infty} k_l P_l(x).
\label{k1}
\end{equation}
The $k_l$ are determined by projecting the kernel $K$ onto the
Legendre polynomials, using
 the
normalization relation 
$\int_{-1}^{1} P_{l'}(x) P_l(x) dx = \frac{2}{2l+1} \delta_{l l'}$.
Therefore
\begin{eqnarray} 
k_l&=&\frac{2l+1}{4 \pi} \int_{-1}^{1} \delta(x-\cos \alpha) P_l(x) dx,\\
&=&\frac{2l+1}{4 \pi}  P_l(\cos \alpha). \label{coeff}
\end{eqnarray}
Next we use the addition theorem for spherical harmonics
 \cite{jackson}
\begin{equation}
P_l({\bf \hat{t}}\cdot {\bf \hat{t}}')=\frac{4 \pi}{2l+1} \sum_{m=-l}^{l}
Y_{lm}^*(\hat{\bf t}') Y_{lm}(\hat{\bf t}), \label{addthm}
\end{equation}
Substituting equations \eqref{addthm} and \eqref{coeff} in equation
\eqref{k1}, we have
\begin{equation} 
K_{\alpha}({\bf \hat{t}},{\bf \hat{t}}')= \sum_{l=0}^{\infty}   P_l(\cos
\alpha)  \sum_{m=-l}^{l}
Y_{lm}^*(\hat{\bf t}' ) Y_{lm}(\hat{\bf t} ). 
\label{ksph}
\end{equation} 
Note that if $\alpha=0$, the kink operator reduces to the identity
because $P_l(1)=1$. 

The probability distribution  $\psi$ just before the bend is
\begin{equation} 
\psi({\bf \hat{t}}',s_o-\epsilon) = \sum_{j=0}^{\infty}
\sum_{k=-j}^{j} \Psi_{jk}(s_o-\epsilon) 
Y_{jk}(\hat{\bf t}').
\label{series2}
\end{equation}
Note that in the case of azimuthal symmetry, the terms with $k \neq 0$
are zero. To determine $\psi$ just after the bend, we substitute the
expressions in equations \eqref{ksph} and \eqref{series2} into the
formula
\begin{equation} 
\psi({\bf \hat{t}},s_o+\epsilon)= \int d {\bf \hat{t}}' \ K_{\alpha}({\bf
  \hat{t}},{\bf \hat{t}}') \ \psi({\bf \hat{t}}',s_o-\epsilon).
\label{trans}
\end{equation}
The expression simplifies by the orthonormality of spherical
harmonics:
\begin{eqnarray} 
\psi({\bf \hat{t}},s_o+\epsilon)&=& \sum_{l=0}^{\infty} P_l(\cos
\alpha)   \sum_{m=-l}^{l}  Y_{lm}(\hat{\bf t} )
\sum_{j=0}^{\infty} \sum_{k=-j}^j \Psi_{jk}(s_o-\epsilon) \int d {\bf
  \hat{t}}' \ Y_{lm}^*(\hat{\bf t}' ) 
 Y_{jk}(\hat{\bf t}' ) \\
&=& \sum_{l=0}^{\infty} P_l(\cos
\alpha)   \sum_{m=-l}^{l}  Y_{lm}(\hat{\bf t} )
\sum_{j=0}^{\infty} \sum_{k=-j}^j \Psi_{jk}(s_o-\epsilon) \delta_{jl}
\delta_{mk} \\ 
&=& \sum_{l=0}^{\infty}\sum_{m=-l}^{l}  P_l(\cos
\alpha) \Psi_{lm}(s_o-\epsilon)  Y_{lm}(\hat{\bf t} ).
\end{eqnarray} 
The transformation can thus be written
$\Psi_{lm}(s_o+\epsilon)=P_l(\cos \alpha) \Psi_{lm}(s_o-\epsilon)$.
The probability distribution just after the kink differs from the
distribution before the kink only in the multiplication of each term
in the series by $P_l(\cos \alpha)$. We can represent the
transformation by a diagonal matrix $W$ such that
\begin{equation}
{\bf
  \Psi}(s_o+\epsilon)=W {\bf \Psi}(s_o-\epsilon).
\end{equation}
Because $\psi$ is azimuthally symmetric (only the $m = 0$ terms appear
in the series expansion), $W$ has entries $W_{l,l}=P_{l-1}(\cos
\alpha)$.

\subsubsection{Distribution of bend angles}

The representation of the bend operator in terms of spherical
harmonics remains simple when the bend contains a distribution of
angles described by $h(u)$ (equation \eqref{bendopdist}).  As above,
we expand $K_h(x)$ in Legendre polynomials, $
K_{h}(x)=\sum k_l P_l(x).  $ The $k_l$ are the projection of $h(x)$
onto Legendre polynomials:
\begin{equation} 
k_l=\frac{2l+1}{4 \pi} \int_{-1}^{1} dx\   h(x)
P_l(x). \label{coeff_dist}
\end{equation}
The calculation is then identical to the case of a single bend angle,
with the result $\Psi_{lm}(s_o+\epsilon)=k_l \Psi_{lm}(s_o-\epsilon)$.
We can represent the transformation by a diagonal matrix $W_h$ such
that
\begin{equation}
{\bf
  \Psi}(s_o+\epsilon)=W_h {\bf \Psi}(s_o-\epsilon).
\end{equation}

\subsection{Force-extension relation with bend}

Once the matrix $W$ (which represents the bend operator in the basis
of spherical harmonics) has been determined, calculation of the
force-extension relation is straightforward. Suppose a single bend
occurs at fractional position $a$ along the chain. The partition
function with a bend is
\begin{equation} 
Z_b={\bf \Psi}^T(s=\ell)e^{(1-a) \ell H} W e^{a \ell H}{\bf \Psi}(s=0),\\
\label{zbend}
\end{equation} 
As before, we let $B=e^H$, denote by $\lambda_*$ the largest
eigenvalue of $B$, and define ${\cal B}=B/ \lambda_*$.  Using $e^{\ell
  H} = \lambda_*^{\ell} {\cal B}^{\ell}$, the logarithm of the
partition function is 
\begin{equation}
  \ln Z_b=  \ell \ln \lambda_* + \ln [{\bf \Psi}^T(s=\ell)\ {\cal
    B}^{(1-a) \ell} W  {\cal B}^{a \ell} \ {\bf \Psi}(s=0)]. 
\label{znewbend}
\end{equation}
As above, the extension is 
$z/L=\ell^{-1}\ \partial \ln Z/\partial f.$

\section{Results}

Here we predict the magnitude of extension change induced by a single
bend, in order to understand when such single-bending events will be
experimentally detectable. We describe how the extension change
induced by a bend depends on  applied force,  bend angle,
contour length, and the position of the bend.

\begin{figure} 
   \begin{center}
      \includegraphics*[width=2.5in]{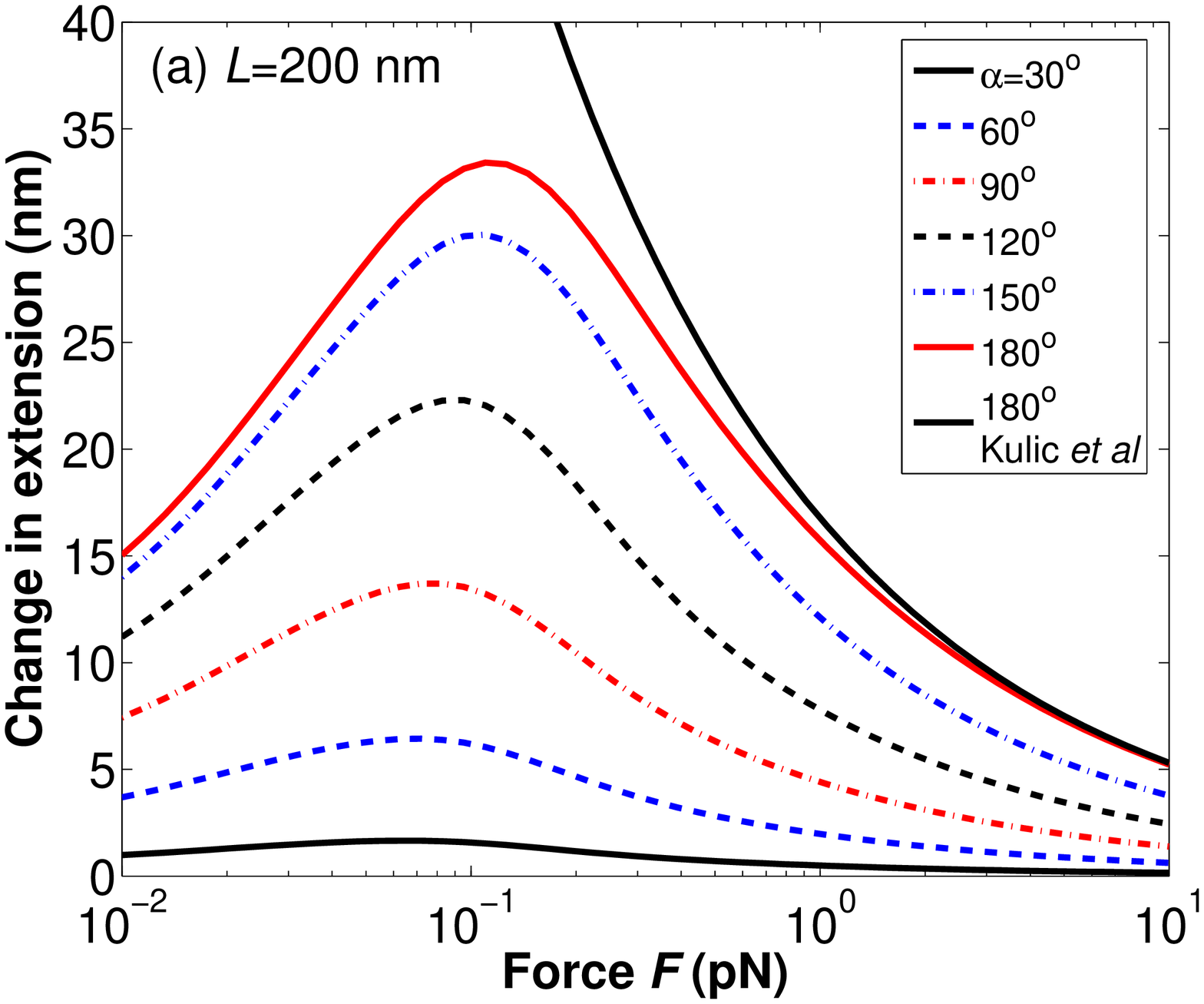}
      \includegraphics*[width=3in]{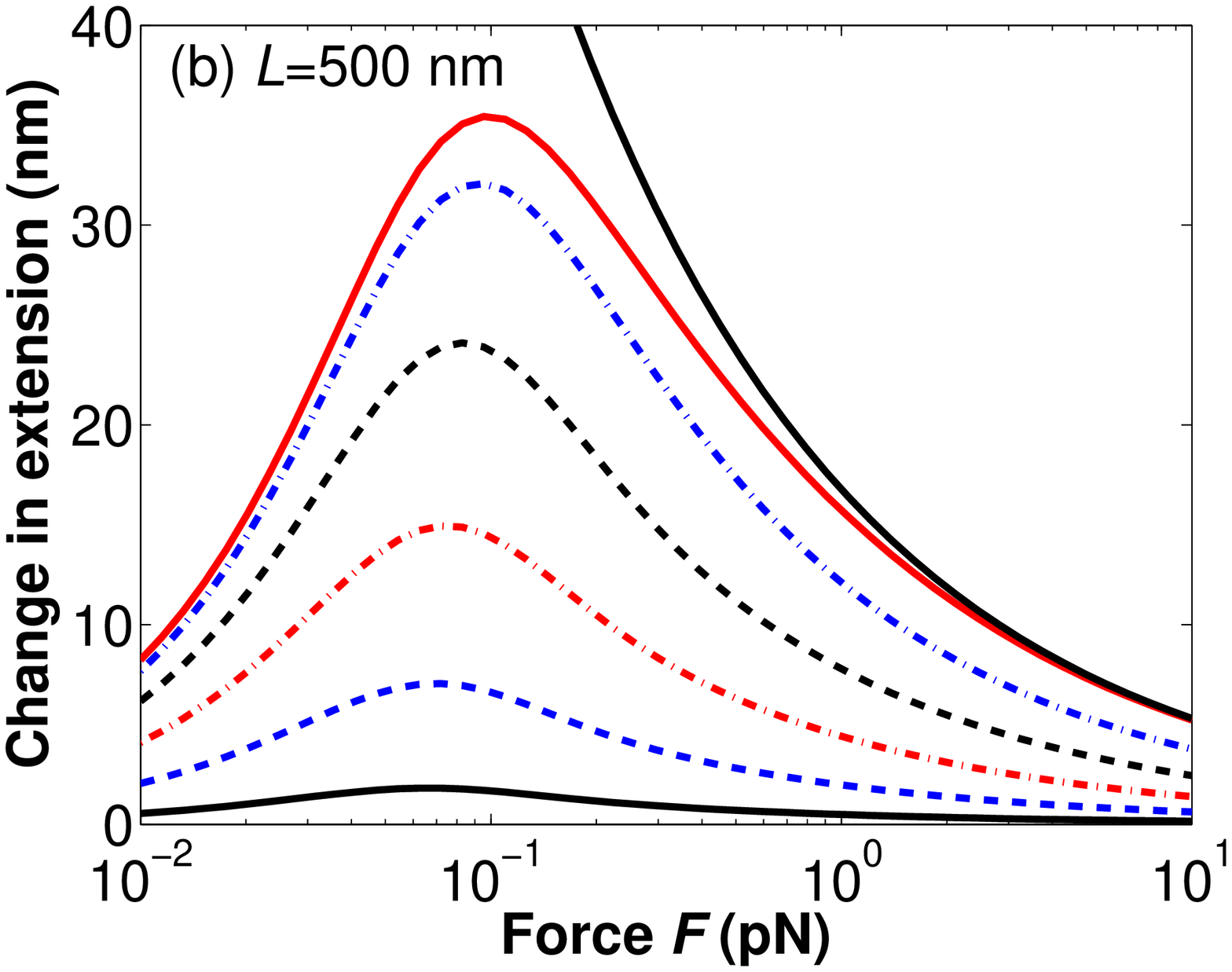}
      \caption{Change in extension due to a bend as a function
        of force, determined by subtracting the extension of the chain
        with the bend from the extension without the bend. The contour
        length is (A) $L$=200 nm and (B) $L$=500 nm. For larger
        contour lengths, the prediction is similar to (B). We assume
        $A=50$ nm, the bend is at the midpoint of the polymer, a bead
        of radius 250 nm is attached to one end of the chain, and
        half-constrained boundary conditions. For the largest bend
        angle, we show for comparison the prediction of Kuli\'c
        \textit{et al.} \cite{kulic05}, which is valid in the
        high-force limit.}
      \label{forcext}
   \end{center} 
\end{figure}

In figure \ref{forcext} we show the change in extension induced by a
bend: the extension of the chain without the bend minus the extension
of the chain with the bend. As expected, the extension change is
larger when the bend angle is larger. In addition, we find that the
change in extension has a maximum near an applied force of 0.1 pN. At
this force, the change in extension due to the bend is a significant
fraction of the persistence length (10-30 nm for $A=50$ nm).

For the largest bend angle, we show for comparison the prediction of
Kuli\'c \textit{et al.} \cite{kulic05}. The Kuli\'c \textit{et al.}
result is valid in the high-force limit, and we find that their
prediction and our result converge as the force becomes large. The
Kuli\'c \textit{et al.} result is valuable because it is a simple
analytical expression. Although our results are obtained numerically,
they are valid over the entire force range.

As the applied force increases, the polymer becomes more stretched and
aligned with the force, decreasing the effect of the bend. For a
classical elastic rod where thermal fluctuations are a weak
perturbation the characteristic propagation length of elastic
deformations is $\sqrt{k_b T\ A /F}$.  Therefore, as the force
increases, the region of the chain experiencing a significant
deflection due to the bend drops.  By this argument, one might expect
that the largest change in extension due to the bend will occur for
the lowest values of the applied force. However, as the force applied
to the ends of the polymer goes to zero, the extension also approaches
zero (on average, there will be no separation of the two ends). In
this case the change in extension due to the bend approaches zero.
The effect of the bend is therefore largest at intermediate force,
where the molecule is extended by the force but not fully extended.

We predicted the change in extension due to the bend with and without
a bead attached to one end of the DNA, and for different values of the
bead radius. In all cases, we predict similar values for the change in
extension due to a bend (not shown).

\begin{figure} 
   \begin{center}
      \includegraphics*[width=2in]{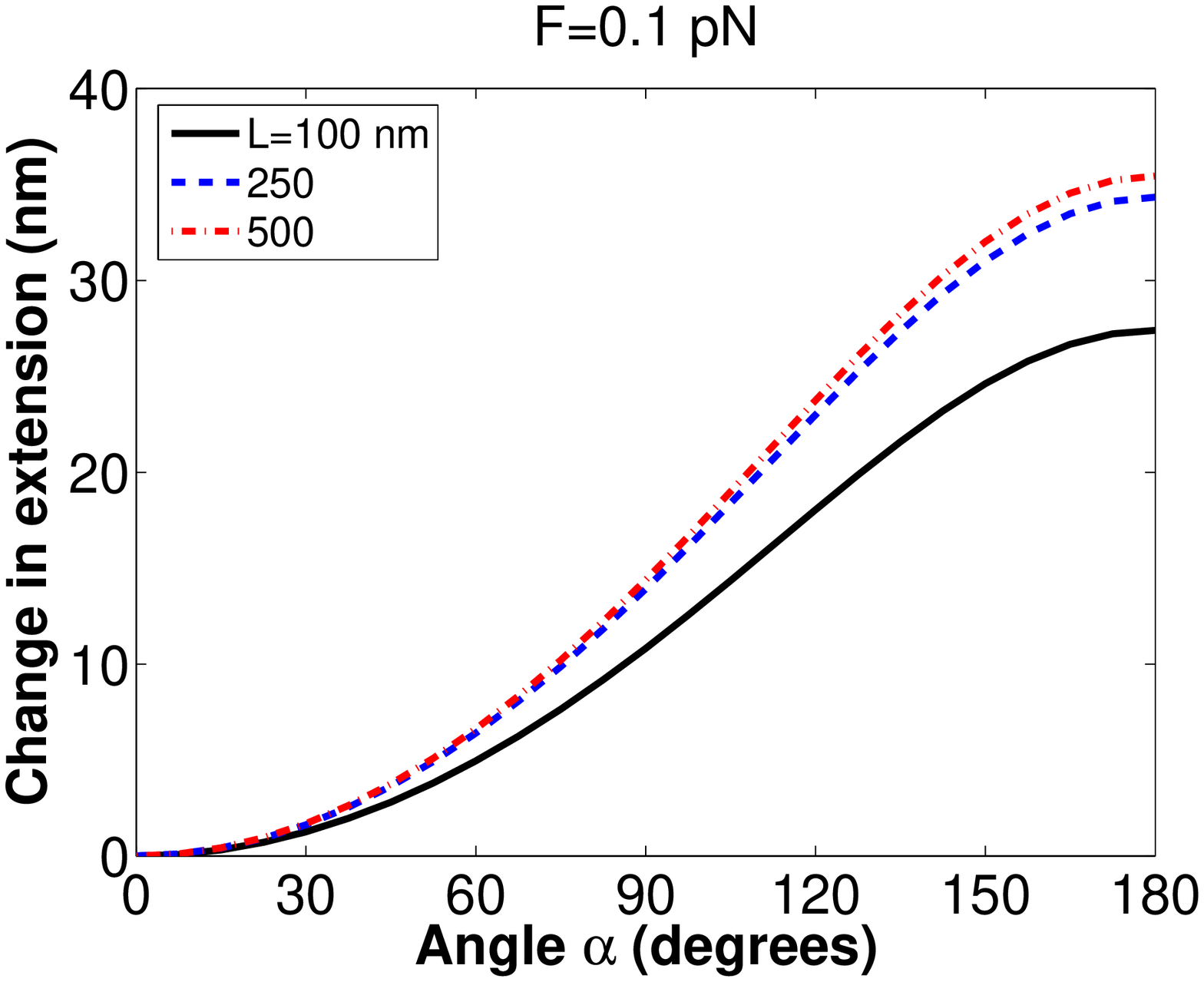}
      \includegraphics*[width=2in]{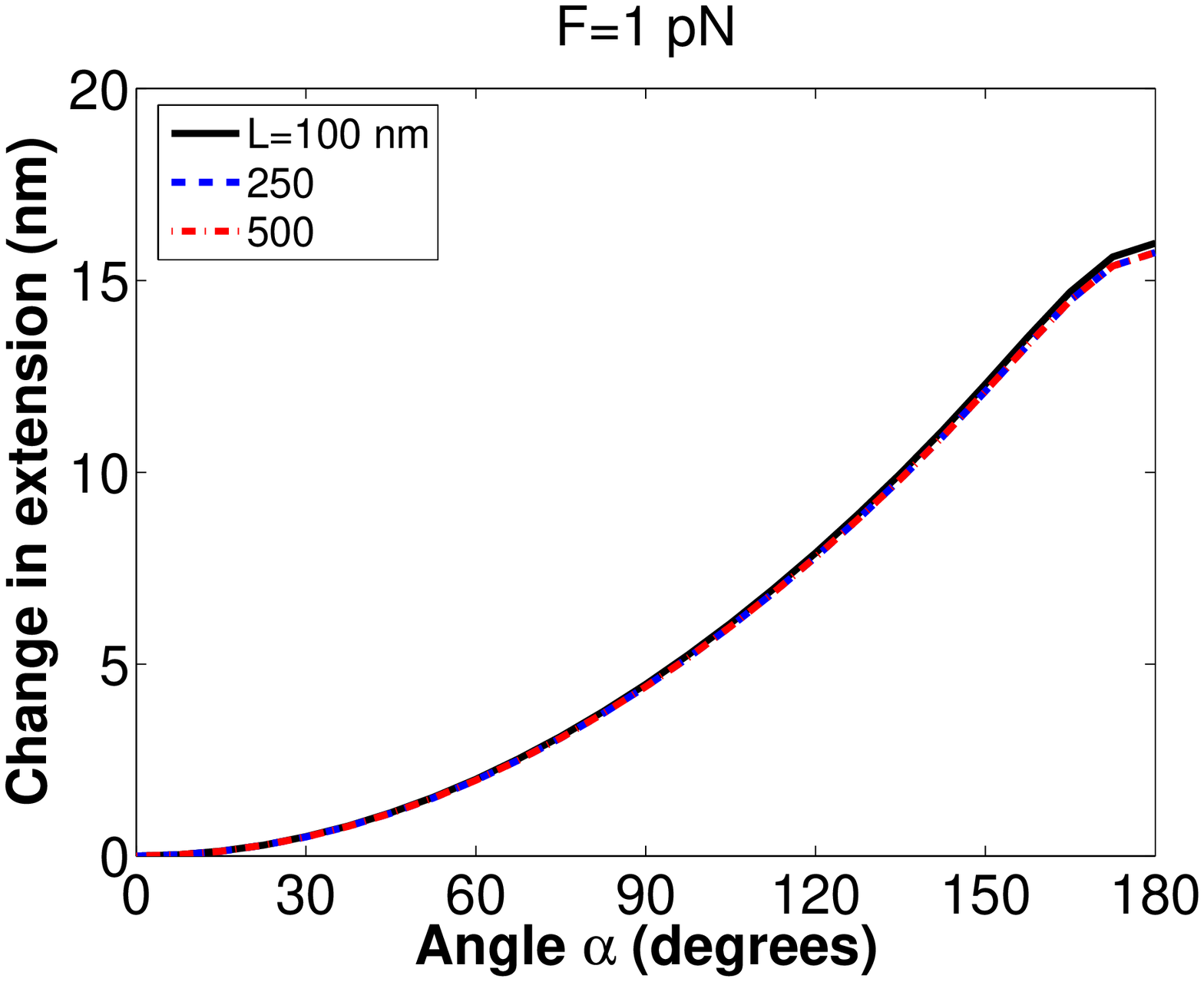}
      \caption{Change in extension due to a bend as a function
        of angle.  (A) F=0.1 pN. (B) F=1 pN. 
        We assume $A=50$ nm, the bend is at the midpoint of the
        polymer, a bead of radius 250 nm is attached to one end of the
        chain, and half-constrained boundary conditions.}
      \label{anglext}
   \end{center} 
\end{figure}

In figure \ref{anglext} we show how the change in extension induced by
the bend varies with bend angle.  The dependence of the extension
change on angle is strong, suggesting that high-resolution experiments
could measure the bend angle by measuring the change in extension due
to a bend. For larger values of the applied force ($F\ge$1 pN), the
result is independent of contour length of the polymer. However at low
force ($F$=0.1 pN), where the change in extension due to a bend is
largest, the results depend on the polymer contour length.

\begin{figure} 
   \begin{center}
      \includegraphics*[width=2in]{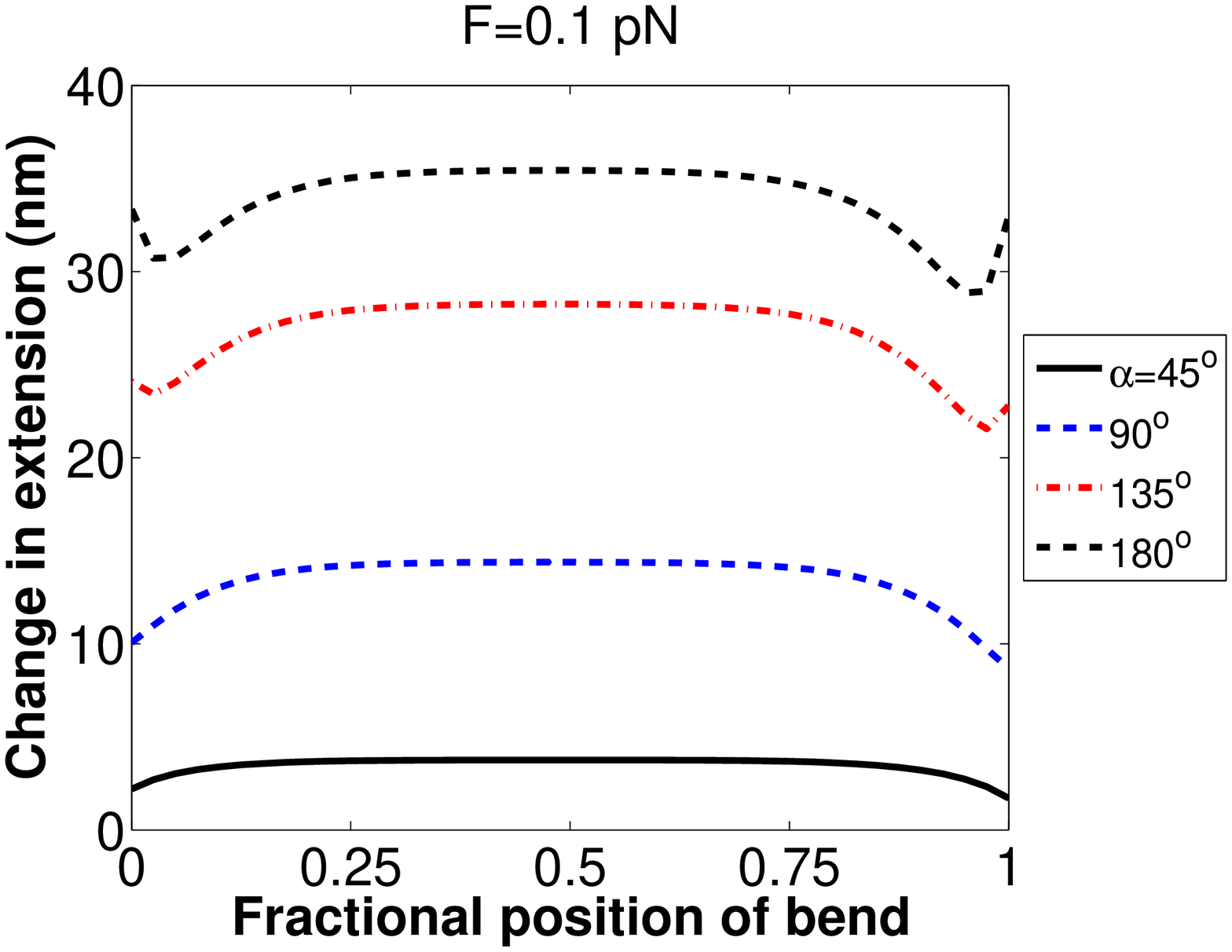}
      \includegraphics*[width=2in]{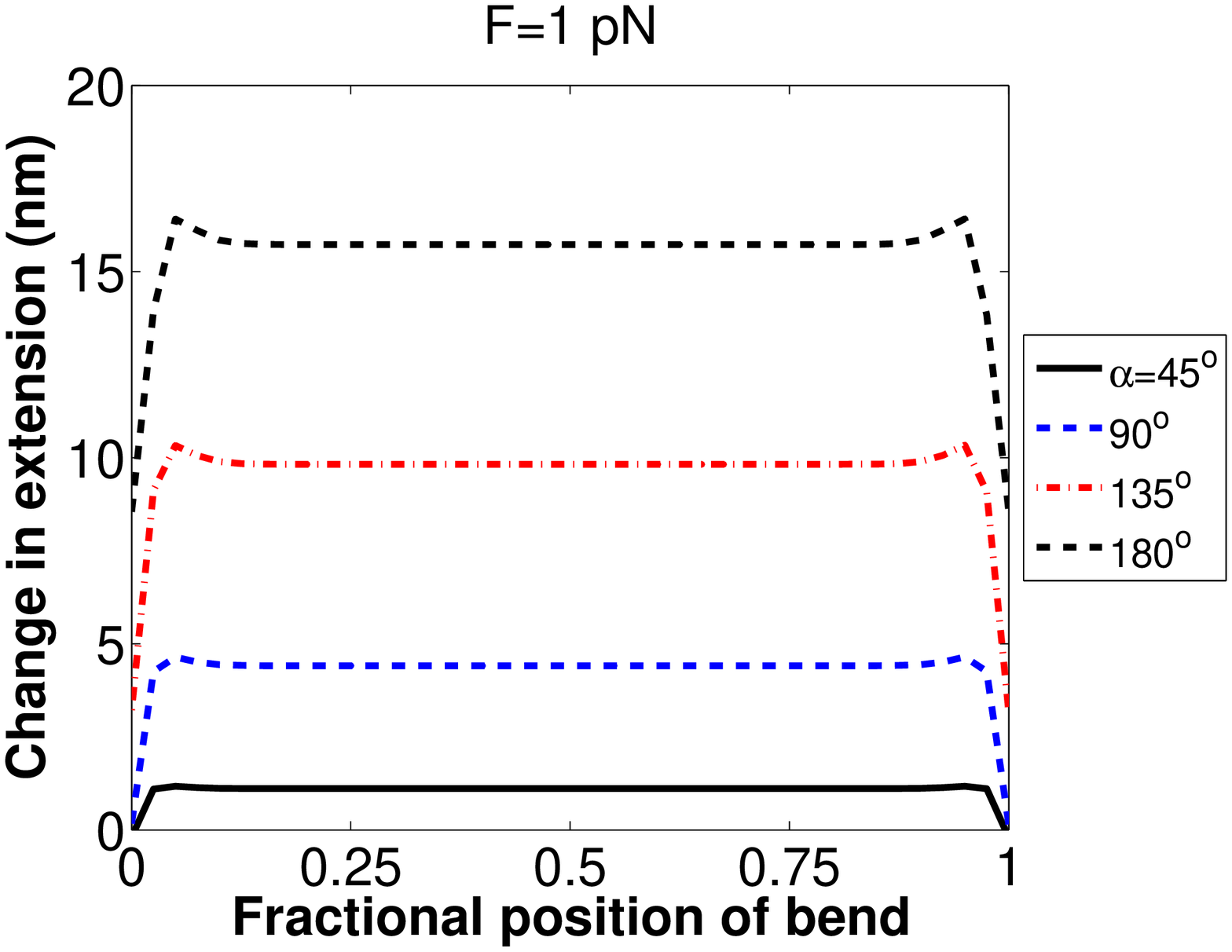}
      \caption{Change in extension due to a bend as a function
        of the position of the bend along the polymer. (A) F=0.1 pN.
        (B) F=1 pN. 
        Note the difference in scale between the two panels.
        We assume $A=50$ nm, $L=$500 nm, a bead of radius 250 nm is
        attached to one end of the chain, and half-constrained
        boundary conditions. }
      \label{bendpos}
   \end{center} 
\end{figure} 

The dependence on the position of the bend is weak, unless the bend is
within a few percent of one end of the polymer (figure
\ref{bendpos}). 
We note that the curves in figure \ref{bendpos} are
not reflection symmetric about the middle of the polymer. This occurs
because we assume one end of the polymer ($s=0$) is attached to a fixed
surface, while the other end of the polymer ($s=1$) is attached to a
bead which can undergo rotational fluctuations. We chose to plot this
case because it is a typical experimental geometry; in the case that
both ends of the polymer experience identical boundary conditions,
then the effects of a bend obey reflection symmetry about the middle
of the polymer.

\section{Discussion}

We have described a theory of DNA elasticity applicable to bent DNA
molecules. The finite worm-like chain model (FWLC) of polymer
elasticity extends the WLC to polymers with $L/A \sim 1-10$
\cite{li05}. The FWLC includes chain-end boundary conditions and
rotational fluctuations of a bead attached to the end of the polymer,
modifications which are important for polymers with contour length a
few times the persistence length.

This work allows predictions of DNA force-extension behavior when a
single bend occurs at a specified point along the chain. When the bend
angle is constant (independent of applied force) the bend operator is
diagonal in the basis of spherical harmonics, allowing straightforward
calculation of the effects of a bend. This mathematical description of
a bend is suitable both for a bend with a single angle and for bends
with a distribution of different bend angles.

We demonstrate that the change in polymer end-to-end extension induced
by the bend can be a significant fraction of the polymer persistence
length: $\Delta z/A \sim 0.2-0.7$ for bend angles of 90-180$^o$, or
$\Delta z \sim 10-35$ nm for dsDNA, which has persistence length of
approximately 50 nm. The change in extension due to the bend is
predicted to show a maximum for applied force around 0.1 pN; for
larger force the polymer conformation becomes highly extended and the
influence of the bend decreases, while for low force the polymer
extension approaches zero, independent of the presence of the bend.

The alterations in polymer extension induced by the bend should be
detectable in high-resolution single-molecule experiments. Since
recent work in single-molecule optical trapping with DNA has
demonstrated a resolution of a few nm\cite{perkin04,nugen04}, DNA
extension changes of 10-35 nm due to a bend should be detectable.
Furthermore, the predicted change in extension strongly depends on the
bend angle, suggesting that high-resolution single-molecule
experiments could directly estimate the angle of a protein-induced
bend.

\section*{Acknowledgements}

We thank Igor Kuli\'c, Tom Perkins, Rob Phillips, and Michael Woodside
for useful discussions, and the Aspen Center for Physics, where part
of this work was done. PCN acknowledges support from NSF grant
DMR-0404674 and the NSF-funded NSEC on Molecular Function at the
Nano/Bio Interface, DMR-0425780. MDB acknowledges support from NSF
NIRT grant PHY-0404286, the Butcher Foundation, and the Alfred P.
Sloan Foundation.  MDB and PCN acknowledge the hospitality of the
Kavli Institute for Theoretical Physics, supported in part by the
National Science Foundation under Grant PHY99-07949.


\begin{thebibliography}{10}
\bibitem{normnote}
Note that for the computation of the force-extension relation, it is not
  necessary to properly normalize the probability distribution, because we are
  computing the derivative of the logarithm of $Z$. Our expressions for the
  probability distribution vectors will neglect the constant normalization
  factor.
\end{thebibliography}

\providecommand{\refin}[1]{\\ \textbf{Referenced in:} #1}

\end{document}